\documentclass[3p,times]{elsarticle}

\usepackage{ecrc}

\usepackage{epstopdf}
\usepackage{wrapfig}

\newcommand\la{\langle}
 \newcommand\ra{\rangle}
 \newcommand\beq{\begin{equation}}
 \newcommand\noi{\noindent}
 \newcommand\eeq{\end{equation}}
 \newcommand\beqn{\begin{eqnarray}}
 \newcommand\eeqn{\end{eqnarray}}
 \newcommand\GeV{{\rm GeV}}

\def\fm{\,\mbox{fm}}
\def\GeV{\,\mbox{GeV}}

\volume{00}

\firstpage{1}

\journalname{Nuclear Physics A}

\runauth{B. Z. Kopeliovich et al.}

\jid{nupha}

\jnltitlelogo{Nuclear Physics A}

\usepackage{graphicx}
\usepackage{amsmath,amssymb}

\begin{document}

\begin{frontmatter}

\title{Charmonium in a hot medium: melting vs absorption}

\author{B.Z. Kopeliovich}
\author{I.K. Potashnikova}
\author{Ivan Schmidt}
\author{M. Siddikov}
\address{Departamento de F\'{\i}sica,
Universidad T\'ecnica Federico Santa Mar\'{\i}a; and\\
Centro Cient\'ifico-Tecnol\'ogico de Valpara\'{\i}so,
Avda. Espa\~na 1680, Valpara\'{\i}so, Chile}

\begin{abstract}
 A charmonium produced  in heavy ion collisions at RHIC and LHC propagates through a dense co-moving matter with a rather high relative momentum, $\la p_T^2\ra=4-10\GeV^2$. In spite of Debye screening of the binding potential, the charmonium survives with a substantial probability, even if the $\bar cc$ potential is completely screened in the hot environment.  In addition, the color-exchange interaction with the medium is another important source of charmonium suppression. Attenuation in a hot medium caused by both effects is evaluated by means of the path integral technique, which requires ability of boosting the binding potential to a moving reference frame. This problem is solved in the approximation of small intrinsic velocities of the charmed quarks.
 
 \end{abstract}

\begin{keyword}
charmonium \sep melting \sep absorption

\end{keyword}

\end{frontmatter}

\section{Introduction}
\label{intro}

We update the theoretical tools for description of propagation and attenuation of a colorless $\bar cc$ dipole in a hot matter created in heavy ion collisions. We employ the path integral technique, where the imaginary part of the potential in the Schr\"odinger equation is related to the effects of interaction with the medium (Sect.~\ref{absorption}). 
The real part of the potential is responsible for the interaction between $c$ and $\bar c$, described by a potential, which has been known so far only in the rest frame of the charmonium. 
A procedure of Lorentz boosting the binding potential, which is required to describe Debye screening for a moving charmonium, is developed (Sect.~\ref{melting}). We evaluate  $J/\psi$ suppression in heavy ion collisions at realistic values of the transport coefficient and found both effects  to be comparable (Sect.~\ref{results}).

\section{Absorption}\label{absorption}

A colorless $\bar cc$ dipole can interact with a target exchanging gluons (quark exchange is suppressed by the OZI rule), turning the $\bar cc$ into a color octet state, which hadronises mainly producing an open charm states, rather than a charmonium. We label such a breakup process as absorption. The cross section is subject to color transparency, namely, it vanishes
as $r_T^2$ for dipoles with a small transverse separation $r_T$ \cite{zkl}. A quantum-mechanical description of the dipole evolution can be
performed with the path-integral method \cite{kz91,kst1}. One has to sum up all possible trajectories of quark and antiquark propagation in order to
incorporate the effects of 
absorption and fluctuations of the dipole separation. 
In the light-cone variables the Green function of the dipole propagating in a medium satisfies
the 2-dimensional Schr\"odinger-type equation~\cite{kz91,kst1} (see detailed derivation in the Appendix of hep-ph/9808378) 
\beq
i\frac{\partial}{\partial l}G\left(l,\vec r_T;0,\vec r_{0}\right) =
\left[2\,\frac{m_c^2-\Delta_{r_T}}{E_\psi}
+U_{\bar cc}\left(r_T,l\right)\right]
G\left(l,\vec r_T;0,\vec r_{0}\right),
\label{318}
\eeq
Here $l$ is the distance covered by the $\bar cc$ dipole
in the rest frame of the medium. The dipole evolves its transverse separation from the initial value $\vec r_0$ 
at $l=0$ up to $\vec r_T$ after propagating distance $l$. We remind that $r_T$ is transverse relative to the charmonium trajectory, which itself is perpendicular to the nuclear collision axis.

The imaginary part of the potential is responsible for absorption,
\beq
{\rm Im}\,U_{\bar cc}(r_T,l)=-{1\over4}\,\hat q(l)\,r_T^2.
\label{350}
\eeq 

The survival probability amplitude of a charmonium produced inside a hot medium is given by
the convolution of the Green function with the initial and final distribution amplitudes,
\beq
S(l) = \frac{\int d^2 r_1 d^2 r_2\Psi_{f}^\dagger(r_2)
G(l,\vec{r_2};0,\vec{r_1})
\Psi_{in}(r_1)}
{\int d^2 r\, \Psi_{f}^\dagger(r)\,\Psi_{in}(r)}
\label{400}
\eeq
The final one $\Psi_{f}(r,\alpha)$ is the charmonium wave function, which is dominated by equal sharing of the light-cone momentum by $c$ and $\bar c$ (see section \ref{melting}).

\section{Debye screening }\label{melting}

The real part of the effective potential  is related to the binding potential, ${\rm Re}\,U(r_T)=V(r_T)$, which is subject to Debye screening within a hot medium. However,
a Lorentz boost of this potential from the $\bar cc$ rest frame is a theoretical challenge, which we deal with below.

\begin{wrapfigure}{r}{0.40\textwidth}
 \vspace{-15pt}
  \begin{center}
    \includegraphics[width=0.40\textwidth]{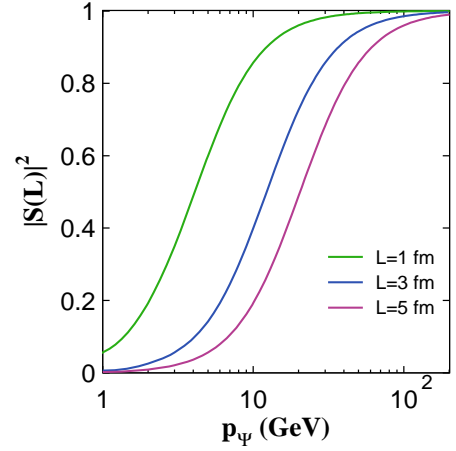}
  \end{center}
  \vspace{-20pt}
  \caption{\label{fig:maximal}(Color online) The attenuation factor (\ref{137}) for $J/\psi$ produced with momentum $p_{\psi}$ off a hot medium. The binding potential is assumed to be completely screened  during in-medium propagation of the $\bar cc$ over the path  $L$}
 \vspace{-15pt}
\end{wrapfigure}

On the contrary to the usual expectation, a moving charmonium is not destructed by Debye color screening \cite{satz}, even if the bound level disappears. Indeed,  
 intuitively is clear that if the binding potential is weakened (for whatever reason), or even completely  eliminated,  for a short time interval $\Delta t\ll1/\omega$ in the charmonium rest frame, the bound state will not be much affected, because the quarks have no time to move away from their orbits. In the medium rest frame, where the charmonium is propagating with momentum $p_\psi\!\equiv\!p_T$,  there are two time scales \cite{brodsky-mueller,kz91}, the coherence time $t_c=E_\psi/(2m_c^2)$, taken for creation of a $\bar cc$ pair, and formation time $t_f=E_\psi/(2m_c\omega)$, needed to form the wave function of the final charmonium.
Here $\omega=(m_{\psi'}-m_{J/\psi})/2$ is the oscillator frequency, characterising the time of circling of $\bar cc$ over the orbit in the bound state. Disappearance of the potential for a short while corresponds in this reference frame  to a path length $L$ in the medium much shorter than the formation length, i.e. $p_\psi\gg 2m_c\omega L\sim 5/v\times L(\fm)$. 
 In such a case the Debye screening effect is expected to vanish, i.e. the medium to become transparent for a high momentum charmonium. This can be demonstrated on an exaggerated example of  a completely terminated binding potential 
inside the medium (maximal Debye screening), which restores back to the vacuum value outside:
${\rm Re}\,U(r)=0$ along the in-medium path length  $l<L$; and
${\rm Re}\,U(r)={\rm Re}\,U_{vac}(r)$  outside the medium, $l>L$.
To clear up this example we simplify the scenario avoiding other sources of attenuation and inessential technical complications. In particular,  we use the oscillator form of the real part of the potential outside the medium and the Gaussian shape  for initial distribution function in (\ref{400}), so both initial and final wave functions have the form $\Psi(r_T)=(\gamma/\sqrt{\pi})\,\exp(-\gamma^2r^2/2)$ with $\gamma^2=m_c^2+p_{\psi}^2/4$ and $\gamma^2=m_c\,\omega/2$ respectively.
Then the Eq.~(\ref{318}) can be solved analytically, and Eq.~(\ref{400}) for a $\bar cc$ propagating with velocity $v=p_{\psi}/E_{\psi}$ over the path length $L$ takes the form \cite{kz91},
\beq
S(L)=-\frac{i\,p_{\psi}}{8L}\,\frac{2m_c^2+m_c\,\omega}{4\pi^2}\,
\left|\int\limits_{-\infty}^\infty dx_1dx_2\,
\exp\left\{-{1\over2}x_1^2\,m_c^2 
-{1\over4}x_2^2\,m_c\omega
-\frac{i\,p_{\psi}}{8L}(x_1-x_2)^2\right\}\right|^2.
\label{137}
\eeq
This suppression  factor is plotted in Fig.~\ref{fig:maximal} vs  $J/\psi$ momentum $p_{\psi}\!\equiv\! p_T$ for $L=1,\ 3,$ and $5\,\fm$.
Obviously, the $J/\psi$ survival probability vanishes towards $p_{\psi}=0$ because $c$ and $\bar c$  fly away with no chance to meet again. Later, however we will see that in reality even at small momenta $p_\psi$ the survival probability is fairly large  due to fast cooling of the medium.
In another limit of high $p_\psi$ the medium becomes fully transparent because the initial tiny dipole size is "frozen" by Lorentz time dilation, and color transparency is at work.

%%%%%%%%%%%%%%%%
%%%%%%%%%%%%%%%%
To proceed further one needs a model for the binding potential, as well as the screening corrections, and we rely on the realistic Cornell form \cite{cornell} with screening corrections calculated on the lattice and parametrized in 
\cite{Karsch:1987pv},
\beq
V_{\bar cc}\left(r,\, T\right) =\frac{\sigma}{\mu(T)}\,
\left(1-e^{-\mu(T)r}\right)-\frac{\beta}{r}\,e^{-\mu(T)r},
\label{280}
\eeq
where $\mu(T) =24\pi^{2}T\sqrt{N_{c}/3+N_{f}/6}/[33\ln\left(19T/\Lambda_{\bar{MS}}\right)]$;
and $\beta\approx0.471$. The bound states in such a potential
are destructed at high temperatures, when the Debye radius $r_{D}\sim1/\mu(T)\lesssim \la r\ra_{\psi}$.

Such a 3-dimensional potential is appropriate for a charmonium  at rest in a hot medium, while data show that
most of charmonia at the LHC have large transverse momenta, $\la p_\psi^2\ra= 7-10\GeV^2$ \cite{alice-psi-pt}.
To perform a Lorentz boost we make use of smallness of the asymmetry in fractional light-cone momenta $x$ of $c$-$\bar c$ quarks, which can be evaluated in the rest frame of the charmonium  \cite{psi-kth},
$
\la\lambda^2\ra\equiv\left\la \left(x-1/2\right)\right\ra=
\la p_L^2 \ra/(4m_c^2)=
\la v_L^2\ra/4,
$
where $p_L$ and $v_L$ are the longitudinal momentum and velocity of the quark. The mean quark (3-dimensional) velocity \cite{psi-kth,hikt1} is $\la v^2\ra\approx 0.2$, so $\la\lambda^2\ra=0.017$ is quite small.
We employ smallness of $\lambda$ solving the Bethe-Salpeter equation for the charmonium wave function and arrive at a rather simple result \cite{klss}.
Introducing a variable $\zeta$ Fourier conjugated to $\lambda$,
\beq
\tilde\Psi_{\bar cc}(\zeta,\vec r_T)=
\frac{1}{2\pi}\int\limits_0^1 dx\,
\Psi_{\bar cc}(x, \vec r_T)\,e^{2im_c\zeta(x-1/2)},
\label{240}
\eeq
a boost-invariant Schr\"odinger equation for the Green function is derived,
which replaces the light-cone form (\ref{318}) \cite{klss},
  \beq
\left[i\frac{\partial}{\partial z_+}+2\,
\frac{\Delta_{r_T}+(\partial/\partial\zeta)^2 -m_c^2}{p_\psi^+}
-U_{\bar cc}\left(\zeta,\vec r_T\right)\right]
G\left(z^+,\zeta,\vec r_T;z_1^+,\zeta_1,\vec r_{1T}\right)=0.
\label{320}
\eeq
Here the real part of the light-cone potential ${\rm Re}\,U(\zeta,\vec r_T)$ is related to the rest frame binding potential $V(r)$ as,
$
{\rm Re}\,U(\zeta,\vec r_T)=V\left(r=\sqrt{\zeta^2+r_T^2}\right)
$.
This solution interpolates between the rest frame, where it reproduces the ordinary Schr\"odinger equation, and the light-cone frame, where it reproduced the Lepage-Brodsky solution \cite{LB}.

\section{Numerical results}\label{results}

The transport coefficient $\hat q$ is related to the medium temperature via  
the equations of state. At large $T\gg T_{c}$, where $T_{c}$
is the critical temperature, $\hat{q}\approx3.6\, T^{3}$, but at lower temperatures $T\lesssim T_{c}$,
this relation is more complicated and is parametrized in \cite{Chen:2010te}.
For the time and coordinate dependence of the transport coefficient $\hat q$
we rely on the popular model \cite{Chen:2010te}.
The maximal value $q_0$ is treated as 
a single free parameter to be adjusted to data for each set of nuclei and each collision energy. This is the main goal of such analyses: to probe the medium properties.

Now we are in a position to solve numerically the equation (\ref{320}) and calculate the FSI suppression
factor for $J/\psi$ produced in nuclear $A$-$B$ collision with impact parameter $b$. 
\beq
S_{J/\psi}^2(b)=
\int\limits_0^{2\pi}\frac{d\phi}{2\pi}
\int \frac{d^2s\,T_A(\vec s)T_B(\vec b-\vec s)}{T_{AB}(b)}
\left|\frac{\int d^2r_1 d^2r_2 d\zeta_1 d\zeta_2\Psi_f^\dagger(\zeta_2,\vec r_2)
G(\infty,\zeta_2,\vec r_2;l_0,\zeta_1,\vec r_1)
\Psi_{in}(\zeta_1,\vec r_1)}
{\int d^2rd\zeta\,\Psi_f^\dagger(\zeta,\vec r)\,\Psi_{in}(\zeta,\vec r)}\right|^2,
\label{330}
\eeq
where $\phi$ is the azimuthal angle between the charmonium trajectory and the reaction plane.

An example shown in Fig.~\ref{results} (left) presents the results for central lead-lead collision
at $q_0=2\GeV^2/\!\fm$, which  was found in \cite{knps} from an analysis of high-$p_T$ light hadrons.   
%%%%%%%%%%%%%%%%
 \begin{figure}[tbh]
{\includegraphics[width=5.2 cm]{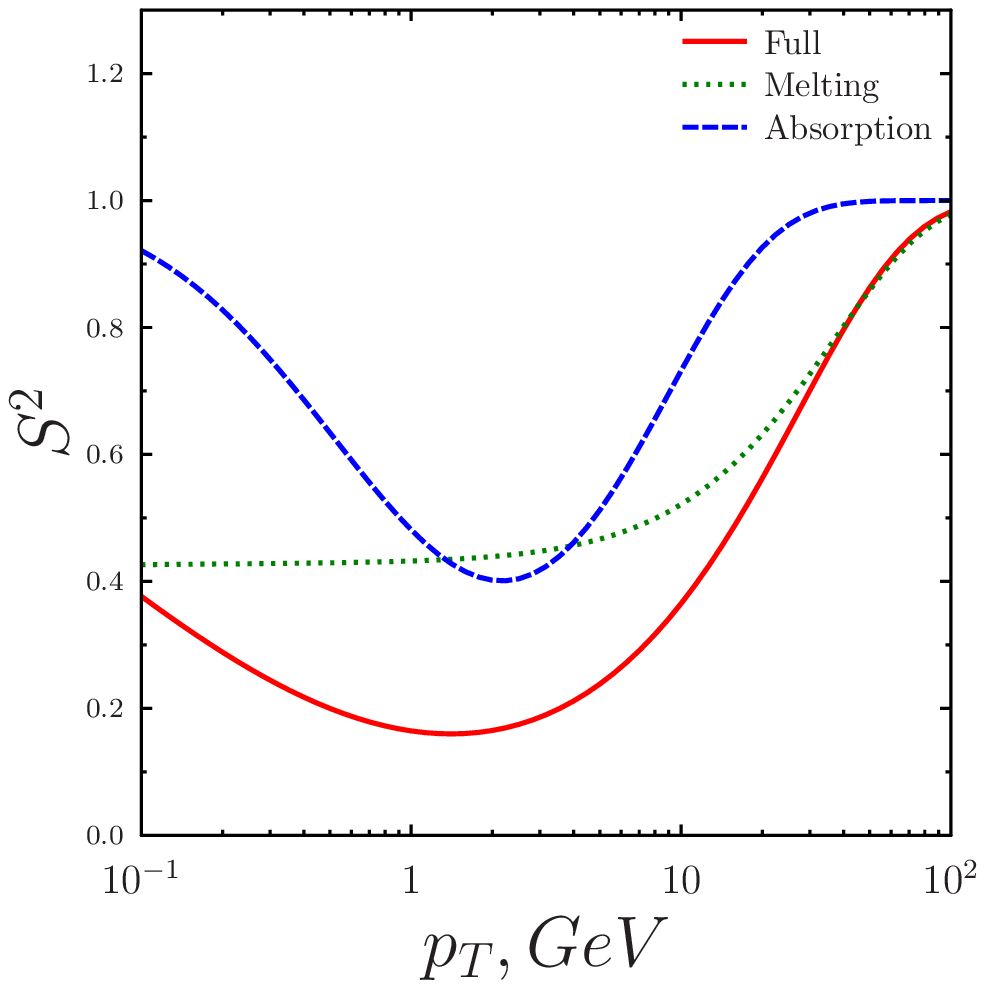}}
\hspace{2mm}
{\includegraphics[width=5.2 cm]{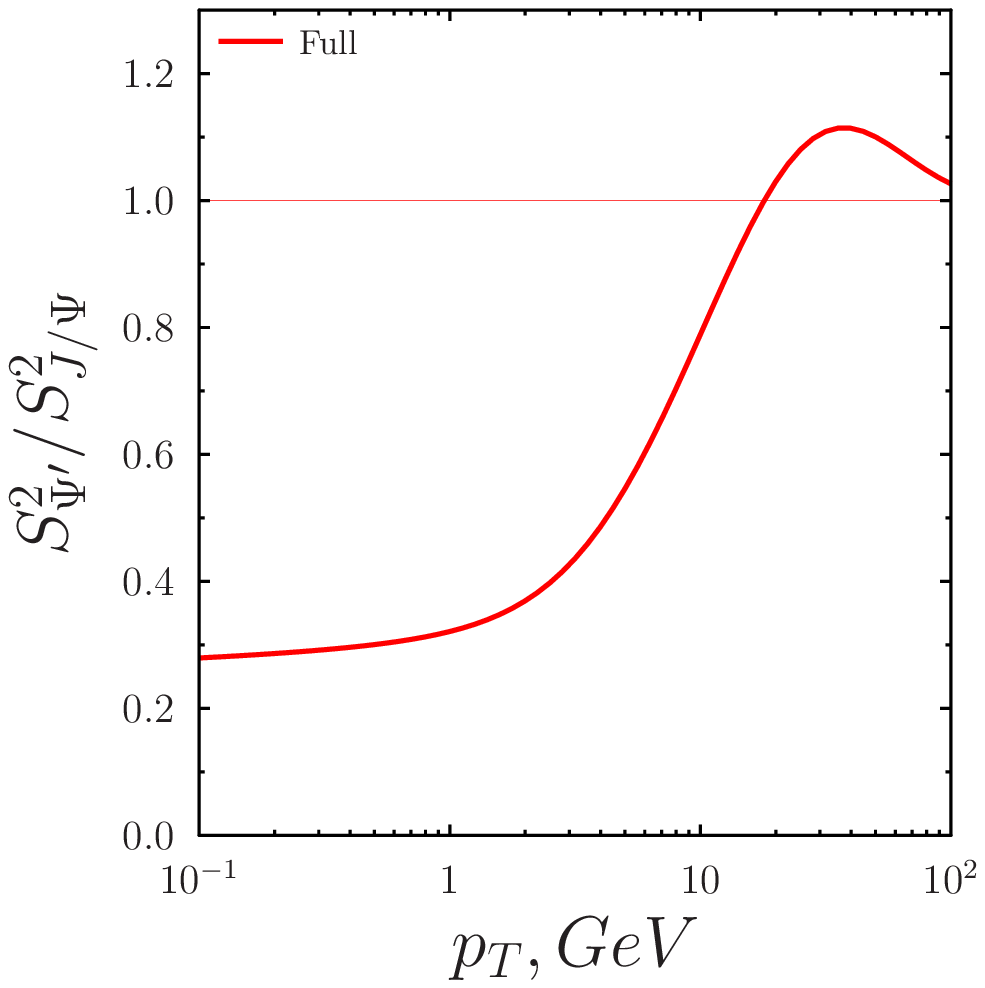}}
\hspace{2mm}
{\includegraphics[width=5.2 cm]{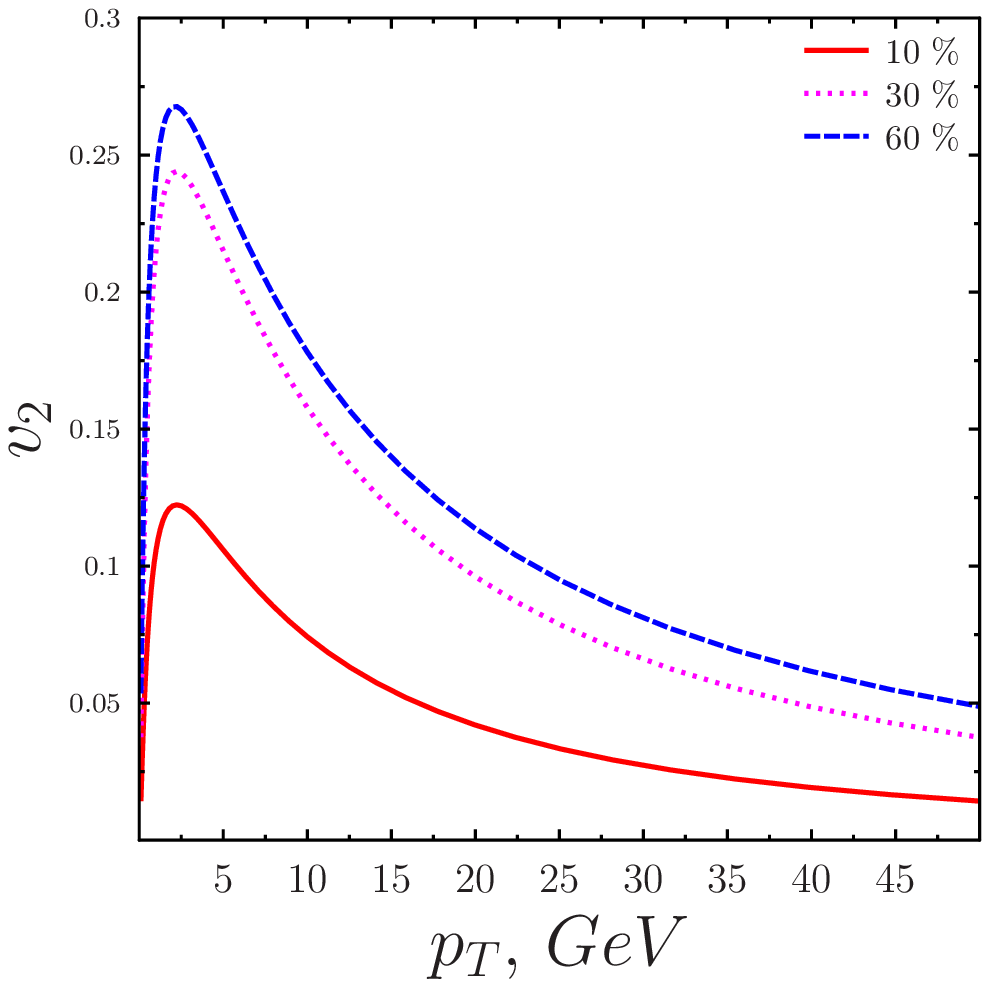}}
\caption{ \label{fig:results} (Color online) {left:} $\left|S(L)\right|^2$ vs $J/\psi$ momentum
$p_\psi\equiv p_T$  for pure melting, absorption, and both effects included, from top to bottom respectively. {\it middle:} Ratio of the suppression factors for $\psi'(2S)$ and $J/\psi$ including the effects of melting and absorption.
{\it right:}  Azimuthal asymmetry parameter $v_2$ for $J/\psi$ production in lead-lead collisions with centralities $10\%,\ 30\%$ and $60\%$.
}
%}
\end{figure}
%%%%%%%%%%%%%%%%
The dashed and dotted curves correspond to net melting (${\rm Im}\,U=0$) or net absorption (${\rm Re}\,U=0$) effects respectively.  The solid curve presents the full solution.

The same effects can be calculated for the radial excitation $\psi'(2S)$, which has a larger radius and should be easier dissolved and absorbed. Indeed, Fig.~\ref{fig:results} (middle)
confirms this. However at higher $p_T\gtrsim 10\GeV$ a bump exceeding unity is observed.
It is related to the complicated structure of the $\psi'$ wave function, which has a node in $r$-dependence.

These results cannot be directly compared with data, because the initial cold matter stage \cite{nontrivial,rhic-lhc} is not considered here. 
Nevertheless,  the asymmetry parameter ${\rm v}_2=\la\cos(2\phi)\ra$
is not affected by the initial stage, so it is worth calculating, averaging $\cos(2\phi)$ with Eq.~(\ref{320}).
The results are depicted in Fig.~\ref
{results} (right) at different centralities.
The magnitude of the effect is close to what was observed for $J/\psi$ at $p_T>7\GeV$ by the CMS experiment recently \cite{cms-v2-psi}. 

Notice that the initial $\bar cc$ dipole was assumed to be colorless,
while with a higher probability a color octet $\bar cc$ may be created,
which then bleaches its color interacting with the medium. This mechanism, enhancing the yield of charmonia at small $p_T$, will be considered elsewhere. 

Summarising, we demonstrated (i) that Debye color screening of the binding $\bar cc$ potential may not destruct a charmonium moving through a hot medium, even if the temperature is extremely high; (ii) An additional strong suppression of a charmonium comes from its color-exchange interactions with the medium; (iii) Both effects are calculated with the path-integral technique, which requires knowledge of a Lorentz boosted Schr\"odinger equation. The corresponding procedure is derived, which interpolated between known limits of the rest and light-cone frames.

\noi
{\bf Acknowledgments:} this work was supported in part
by Fondecyt grants 1130543, 1130549, 1100287, 
and 1140377.


\begin{thebibliography}{00}

\bibitem{zkl}
  B.~Z.~Kopeliovich, L.~I.~Lapidus and A.~B.~Zamolodchikov,
  %``Dynamics Of Color In Hadron Diffraction On Nuclei,''
  JETP Lett.\  {\bf 33}, 595 (1981)
  [Pisma Zh.\ Eksp.\ Teor.\ Fiz.\  {\bf 33}, 612 (1981)].

\bibitem{kz91} 
  B.~Z.~Kopeliovich and B.~G.~Zakharov,
  %``Quantum effects and color transparency in charmonium photoproduction on nuclei,''
  Phys.\ Rev.\ D {\bf 44}, 3466 (1991).
  
  \bibitem{kst1}
B.~Z.~Kopeliovich, A.~V.~Tarasov and A.~Sch\"afer,
  %``Bremsstrahlung of a quark propagating through a nucleus,''
  Phys.\ Rev.\ C {\bf 59}, 1609 (1999)
  [hep-ph/9808378].
  
  \bibitem{brodsky-mueller} 
  S.~J.~Brodsky and A.~H.~Mueller,
  %``Using Nuclei to Probe Hadronization in QCD,''
  Phys.\ Lett.\ B {\bf 206}, 685 (1988).
  
 \bibitem{satz} T. Matsui and H. Satz, Phys. Lett. B \textbf{178},
 416 (1986). 
 
\bibitem{alice-psi-pt} 
  B.~Abelev {\it et al.}  [ALICE Collaboration],
  %``Inclusive $J/\psi$ production in $pp$ collisions at $\sqrt{s} = 2.76$ TeV,''
  Phys.\ Lett.\ B {\bf 718}, 295 (2012)
  [arXiv:1203.3641 [hep-ex]].
  
\bibitem{cornell}
E. Eichten, K. Gottfried, T. Konoshita, K. D. Lane, and T.-M.
Yan, Phys. Rev. D {\bf 17}, 3090 (1978); {\bf 21}, 203 (1980).
    
\bibitem{Karsch:1987pv}F.~Karsch, M.~T.~Mehr and H.~Satz, Z.~Phys.~C
 \textbf{37}, 617 (1988).
 
\bibitem{psi-kth} 
  B.~Kopeliovich, A.~Tarasov and J.~H\"ufner,
  %``Coherence phenomena in charmonium production off nuclei at the energies of RHIC and LHC,''
  Nucl.\ Phys.\ A {\bf 696}, 669 (2001)
  [hep-ph/0104256].

\bibitem{hikt1} 
  J.~H\"ufner, Y.~.P.~Ivanov, B.~Z.~Kopeliovich and A.~V.~Tarasov,
  %``Photoproduction of charmonia and total charmonium proton cross-sections,''
  Phys.\ Rev.\ D {\bf 62}, 094022 (2000)
  [hep-ph/0007111].
  
  \bibitem{klss} B. Z. Kopeliovich, E. M. Levin, I. Schmidt and M. Siddikov,
  paper in preparation.
  
\bibitem{LB} 
  G.~P.~Lepage and S.~J.~Brodsky,
  %``Exclusive Processes in Perturbative Quantum Chromodynamics,''
  Phys.\ Rev.\ D {\bf 22}, 2157 (1980).

   \bibitem{Chen:2010te}X.-F. Chen, C. Greiner , E. Wang , X.-N. Wang, Z. Xu, Phys.Rev.\textbf{ C81} (2010) 064908 {[}arXiv:1002.1165 {[}nucl-th{]}{]}.
 
\bibitem{knps} 
  B.~Z.~Kopeliovich, J.~Nemchik, I.~K.~Potashnikova and I.~Schmidt,
  %``Quenching of high-pT hadrons: Energy Loss vs Color Transparency,''
  Phys.\ Rev.\ C {\bf 86}, 054904 (2012)
  [arXiv:1208.4951 [hep-ph]].

\bibitem{nontrivial} 
  B.~Z.~Kopeliovich, I.~K.~Potashnikova, H.~J.~Pirner and I.~Schmidt,
  %``Heavy quarkonium production: Nontrivial transition from pA to AA collisions,''
  Phys.\ Rev.\ C {\bf 83}, 014912 (2011)
  [arXiv:1008.4272 [hep-ph]].

\bibitem{rhic-lhc} 
  B.~Z.~Kopeliovich, I.~K.~Potashnikova and I.~Schmidt,
  %``Nuclear suppression of J/Psi: from RHIC to the LHC,''
  Nucl.\ Phys.\ A {\bf 864}, 203 (2011)
  [arXiv:1012.5648 [hep-ph]].

\bibitem{cms-v2-psi} Dong Ho Moon (for the CMS Collaboration),
a talk at Quark Matter 2014, Darmstadt, May 19-24, 2014.

\end{thebibliography}
\end{document}